\documentstyle[12pt,psfig,paspconf]{article}
\begin{document}

{\large\bf{A long-duration flare in
the X-ray/EUV selected chromospherically active binary 2RE~J0743+224}}

{\it{ D. Montes$^{1,2}$, L.W. Ramsey$^{1}$
}}

$^1$ {The Pennsylvania State University,
Department of Astronomy and Astrophysics,
525 Davey Laboratory, University Park, PA 16802, USA}\\
$^2$ {Departamento de Astrof\'{\i}sica,
Facultad de F\'{\i}sicas,
Universidad Complutense de Madrid, E-28040 Madrid, Spain}

\vspace*{0.6cm}

To be published  in ASP Conf. Ser., Solar and Stellar Activity:
Similarities and Differences
(meeting dedicated to Brendan Byrne, Armagh 2-4th September 1998)
C.J. Butler and J.G. Doyle, eds

\vspace*{2.5cm}

\baselineskip=0.6truecm

\large

\hrule
\vspace{0.2cm}
\begin{center}
{\huge\bf Abstract}
\end{center}

2RE~J0743+224  (BD +23 1799) is a chromospherically active star
selected by X-rays and EUV emission detected in the
Einstein Slew Survey and ROSAT Wide Field Camara (WFC) all sky survey,
and classified as single-lined spectroscopic binary by (Jeffries et al. 1995).
We present here high resolution echelle spectroscopic observations
of this binary, obtained
during a 10 night run 12-21 January 1998 using the 2.1m telescope at
McDonald Observatory.
These observations reveal it is a double-lined spectroscopic binary.

A dramatic increase in the chromospheric emissions
(H$\alpha$ and Ca~{\sc ii} IRT lines)
is detected during the observations.
Several arguments favor the interpretation of this behavior as
an unusual long-duration flare.
First the temporal evolution of the event is similar to the observed in other
solar and stellar flares, with an initial impulsive phase
characterized by a strong increase in the chromospheric lines
(the H$\alpha$ EW change in a factor of 5 in only one day)
and thereafter, the line
emission decreased gradually over several days.
Second, a broad component in the H$\alpha$ line profile is observed just
at the beginning of the event.
Third, the detection of the He~{\sc i} D$_{3}$ in emission and
a filled-in He~{\sc i} $\lambda$6678~\AA.

We detect a Li~{\sc i} $\lambda$6708~\AA\ line enhancement
which is clearly related with the temporal evolution of the flare.
The maximum Li~{\sc i} enhancement  occurs just after
the maximum chromospheric emission observed in the flare.
We suggest that this Li~{\sc i} is produced by spallation reactions
in the flare.
This is the first time  that such Li~{\sc i} enhancement associate
with a stellar flare is reported, and probably the long-duration of
this flare  is a key factor for this detection.

\vspace{0.4cm}
\hrule

\newpage
\hrule
\vspace{0.2cm}
\begin{center}
{\huge\bf Introduction}
\end{center}

2RE~J0743+224 (BD +23 1799, SAO 79647)  is a EUV source detected during the
ROSAT Wide Field Camera (WFC) all sky survey
(Pound et al. 1993; Pye et al. 1995).
The WFC optical identification program (Mason et al. 1995)
identified it as a chromospherically active star with
EW(Ca~{\sc ii} H) = 4.1 and EW(H$\alpha$) = 1.8.
The optical spectroscopic observations presented by Jeffries et al. (1995)
confirmed it as a chromospherically active with strong H$\alpha$ emission
(EW(H$\alpha$) = 1.4). These authors found it to be a single-lined spectroscopic binary
with spectral type K0 and a rotational velocity, {\it v}sin{\it i}, of 13 km s$^{-1}$.
No orbital  solution is reported and
only a lower limit of 3 days to the period is given.
This star was also detected in X-rays by the
Einstein Slew Survey (Elvis et al. 1992) receiving the name 1ES~0740+22.8
and confirmed as chromospherically active by Schachter et al. (1996)
who give an estimated spectral type K0III/IV and a measured
{\it v}sin{\it i} of 17.0  km s$^{-1}$.

\vspace{0.3cm}

We present here high resolution (0.16~\AA)
echelle spectroscopic observations
of this chromospherically active binary (hereafter CAB), obtained
during a 10 night run 12-21 January 1998 using the 2.1m telescope at
McDonald Observatory and the Sandiford Cassegrain Echelle Spectrograph.
These observations reveal it to be a double-lined spectroscopic binary (SB2)
with a orbital period of 10 days.
The wavelength position of
the lines of both components are marked in Fig.~3.
We have determined the chromospheric contribution in the H$\alpha$ and Ca~{\sc ii} IRT lines
using the spectral subtraction technique
(Huenemoerder \& Ramsey 1987; Montes et al. 1997).
The synthesized spectrum was constructed using the program STARMOD
developed at Penn State (Barden 1985).
The best fits to the 2RE~J0743+224 spectra were obtained using
K1~III standard star for the primary and a K5~V for the secondary,
with a relative contribution to the continuum of 0.85 /0.15.
In Fig.~1a, b we plotted the observed spectra together with the K1~III reference star in the left panel
and the subtracted spectra in the right panel.

\vspace{0.5cm}
\hrule

\newpage
\hrule
\vspace{0.2cm}
\begin{center}
{\huge\bf A long-duration flare}
\end{center}

A dramatic increase in the chromospheric emissions
(H$\alpha$ and Ca~{\sc ii} IRT lines)
is detected during the observations (see Fig.~1a, b).
The increase of the emission start the 3th night (1998 January 15)
reach it maximum at the 5th nigh and at the end of the observations
(1998 January 22) the chromospheric lines was not yet recovered the
quiescent value, being the total duration of the event large than 8 days.
Several arguments favor the interpretation of this behavior as
an unusual long-duration flare.

{\bf - 1)} The temporal evolution of the event
is similar to the observed in other
solar and stellar flares, with an initial impulsive phase
characterized by a strong increase in the chromospheric lines.
The H$\alpha$ emission EW
increases by a factor
of $\approx$5 in only one day (from the 2nd to the 3rd night)
and by a factor of 7 at  the maximum.
After this the emission decreases slowly  (gradual phase) until the
end of the observations. 
The time evolution of the EW(H$\alpha$) during the flare is
displayed in Fig.~2.

{\bf - 2)} A broad component in the H$\alpha$ line profile is observed just
at the beginning of the event.
A two Gaussian components fit to the subtracted spectra is displayed in
the right panel of Fig.~1.

{\bf - 3)} The detection of the He~{\sc i} D$_{3}$ in emission and
a filled-in He~{\sc i} $\lambda$6678~\AA\
as have been observed in other solar and stellar flares
(Zirin 1988; Huenemoerder \& Ramsey 1987;
Montes et al.~1996; 1997; 1998)

\vspace{0.2cm}

This is an unusual long-duration flare and very different from the largest
flares observed in the Sun ($\approx$ hours).
However, long-lasting (2 to 9 days) flares have also been observed in other
CAB as II Peg 
(Berdyugina et al. 1998a), AR Lac (Ottmann \& Schmitt 1994),
HK Lac (Catalano \& Frasca 1994),
HR 5110 (Graffagnino et al. 1995),
CF Tuc (K\"urster \& Schmitt 1996), and
HU Vir (Endl et al. 1997);
and in FK Com-type stars as YY Men (Cutispoto et al. 1992).

\vspace{0.2cm}

The analysis of the TiO 7055\AA\ (O'Neal et al. 1996) band indicates
that a fraction of the stellar surface is covered by starspots.
Strong changes are observed in this band deph from
night to night in the same way that the chromospheric lines change.
This suggest that
the chromospheric region responsible for the flare is
spatially coincident with photospheric cool spot(s)
and could be
produced, as have been observed in the largest solar flares,
by the interaction of new emerging magnetic flux with old magnetic
structures associated to the spot groups.

\vspace{0.4cm}
\hrule

\newpage
\hrule
\vspace{0.2cm}
\begin{center}
{\huge\bf Li~{\sc i} enhancement during the flare}
\end{center}

The Li~{\sc i} $\lambda$6708 \AA\ absorption feature 
is clearly observed in
our spectra (Fig.~3) with a mean EW of 130~m\AA.
The line  appears centered at the
wavelength-position corresponding to the primary component
with no evidence for a contribution from the secondary. 
A careful analysis of the Li~{\sc i} line indicates that
the line profile, EW, and intensity, I, are changing during the observations.
The measured EW and I are given in Table~1
and plotted in Fig.~4,
where we can see that the increase of Li~{\sc i} line
follows the temporal evolution of the flare.
The maximum Li~{\sc i} enhancement (40\% in EW) occurs just after
the maximum chromospheric emission observed in the flare.
In order to test if these variations are real we have also measured the
EW of other photospheric lines.
We have selected several isolated lines in the same spectral order
as the Li~{\sc i} line; Fe~{\sc i} 6710.3~\AA\ and Fe~{\sc i} 6703.6~\AA\
close and similar in strength to the Li~{\sc i} line and the more intense
lines Fe~{\sc i} 6663.4~\AA\ and Ca~{\sc i} 6717.7~\AA\ (see Fig.~3). 
Other intense lines included in different spectral order were also measured.
All are neutral lines and, as the Li~{\sc i} line, the EW should be enhanced 
in a different way depending of their
excitational potential when the the temperature decrease.
These lines are listed in Table~2 together with their
excitational potential and the mean EW, the standard deviation, and the
peak to peak variation (EW$_{\rm max}$ - EW$_{\rm min}$).
The measured EW for each night is plotted in Fig~4.
As can be seen the variations in these lines are very small and
contrary to the Li~{\sc i} line any correlation with the temporal evolution
of the flare is observed.
The peak to peak variation in the Li~{\sc i} line is a factor 3 larger
than in the other lines.
Furthermore, no clear systematic behaviour is observed
with different excitation potentials.

The Li~{\sc i} line variations can be due to different causes, including
line blends with other close lines or lines for the secondary component
in the SB2 system, variations related with starspots and faculae
in the stellar surface, and Li formation by spallation reactions
during the flare.
In the following we discuss each of these possibilities to find
the most plausible hypothesis.

\vspace{0.4cm}
\hrule

\newpage
\hrule
\vspace{0.2cm}
\begin{center}
{\huge\it Line blends?}
\end{center}

One source of variability of the
Li~{\sc i} $\lambda$6708 \AA\ line could be the fact that
it is blended with TiO bands at
6707.29, 6707.92, and 6708.16 \AA\ and CN bands at 6707.64 \AA\
and that these molecular bands become stronger at the lower temperatures
of the starspots.
However the calculations of the Li~{\sc i} abundance in sunspots
that take into account these bands
(Engvold et al. 1970; Ritzenhoff et al. 1997)
concluded that the molecular blend is of low importance.
Thus it seems reasonable assume that this effect in the stellar spectra
is negligible.

At this spectral resolution the Li~{\sc i} line is blended with
the nearby Fe~{\sc i} $\lambda$6707.41~\AA\ line, 
the EW of this line is normally
subtracted from the measured Li~{\sc i} + Fe~{\sc i} in order to obtain
the real value of the Li~{\sc i}  EW.
This Fe~{\sc i} is clearly seen in the spectra of the inactive and Li free star
HR 5340 (K1 III) which we use in the spectral subtraction.
The mean EW measured in these spectra is 20 (m\AA),
which is much smaller than the observed Li~{\sc i} EW.
Furthermore, taking into account that the other photospheric lines
we have measured did not show significant variations, we conclude that this
line will not produce any variation in the measured Li~{\sc i}  EW.

Due to the SB2 nature of this binary in some orbital phases the lines of
the primary component could be blended with the lines from the secondary.
This is clearly seen in the spectrum from the 5th night (which is very
close the conjuntion). The EW of the strongest lines measured during this
night are noticeably larger than in the rest of the nights due to the
contribution (about 15\%) to the EW from the secondary.
We have corrected the EW for this effect by subtracting the EW
of the lines in the secondary measured at other orbital phases where the
lines
are not blended. These corrected values are what we have plotted in
Fig.~4.
However, for weak lines with EW similar to the Li~{\sc i} line
the contribution of the secondary seems to be negligible and this effect is
not observed.

In conclusion the observed Li~{\sc i} line variations do not seem to be
due to any kind of line blends.

\vspace{0.4cm}
\hrule

\newpage
\hrule
\vspace{0.2cm}
\begin{center}
{\huge\it Starspots and faculae?}
\end{center}

The Li~{\sc i} line variations could be related with possible
cold spots and faculae in the stellar surface 
(see the review by Fekel 1996).
Since this line is very temperature sensitive,  
the EW should be enhanced in dark
spots but reduced in the bright facular regions as shown by solar observations
(Giampapa 1984).
While Giampapa (1984) suggests this can substantially alter the EW in stellar
spectra, other authors find this is not the case.
No detection of Li~{\sc i} EW variations in six active dwarfs 
have been reported by Boesgaard (1991).
The calculations of Soderblom et al. (1993) indicate that the effect
is only significant when the fraction of the surface covered by spots is
very high (see also Stuik et al. 1997).
Pallavicini et al. (1993) show by means of spectral synthesis simulations
that the effects may be less pronounced than that suggested by Giampapa (1984)
and found no evidence that changes in the EW is
correlated with the photometric variability due to starspots in four active
stars.
The simulations done by Barrado (1996) also indicated smaller
changes in the EW and even, in certain cases, the presence of faculae can
cancel these changes.
By application of the Doppler imaging technique Hussain et al. (1997)
found no evidence for the Li~{\sc i} abundance being
enhanced or depleted in starspots.

Until now    
significant variations in the Li~{\sc i} EW have been found only in
some stars with very high Li~{\sc i} abundances such as
pre-main sequence stars (Patterer et al. 1993;
Fern\'{a}ndez \& Miranda 1998; Neuh\"auser et al. 1998)
and other young and very active stars
(Robinson et al. 1986; Jeffries et al. 1994; Soderblom et al. 1996).
Large EW variation has been observed at
larger amplitude in V band in V410 Tau 
(peak to peak variability of 0.12~\AA\  when the amplitude in V was 0.6~mag)
(Fern\'{a}ndez \& Miranda 1998). 
This result is confirmed on the young star Par~1724 by
Neuh\"auser et al. (1998).
However, in CAB little or no variations
have been previously reported (Pallavicini et al. 1993)
and in recent observations (Berdyugina et al. 1998b)
of the CAB
II Peg, which exhibits high V band variations and spot filling factors,
very small Li~{\sc i} EW variations (10~m\AA), not quite correlated
with quasi-simultaneous photometric observations, have been found.

The Li~{\sc i} EW variations that we observe are clearly larger than
those reported in other CAB with similar activity levels
and Li~{\sc i} abundance.
In other stars that exhibit large Li~{\sc i} EW variations
other photospheric lines exhibit similar EW variations
(Fern\'{a}ndez \& Miranda 1998),
contrary to the behaviour we observe in this star.
Taking into account all these facts, the starspots, 
on the surface of 2RE~J0743+224, 
not seem to be the primary cause
of the  Li~{\sc i} line variation we observed.

\vspace{0.4cm}
\hrule

\newpage
\hrule
\vspace{0.2cm}
\begin{center}
{\huge\it Spallation reactions?}
\end{center}

The formation of Li by low energy spallation reactions in stellar flares
was originally considered by
Fowler et al. (1955), Canal (1974), Canal et al. (1975).
The possibility of detecting Li~{\sc i} abundance inhomogeneities
resulting from spallation reactions in the solar photosphere
have been discussed by Hultqvist (1974, 1977)
and evidence for such Li formation have been found through the
deexcitation line Li(478keV) resulting from He-He reactions, which
has been detected by Gamma-ray spectral observations  of solar flares
with OSO-7 (Chupp et al. 1973), SMM (Murphy et al. 1990) and
Yohkoh (Yoshimori et al. 1994; Kotov et al. 1996).
The recent calculation of Li production in solar flares (Livshits 1997)
agree with Gamma line observations and suggest that enhancement of Li,
especially in the intensity of the Li~{\sc i} $\lambda$6708 line, 
should be observed in the
Sun and other active stars.
Evidence for a Li enhancement at one umbral position, during a solar flare
is reported by Livingston et al. (1997).

In other stars (including the UV Ceti flares stars)
no evidence of production of Li by nuclear reactions
have been found until now.
The possibility of Li production have been discussed only
in terms of the energy required (Ryter et al. 1970; Karpen \& Worden 1979)
or as a possibility to explain the high Li abundances observed in CAB
(Pallavicini et al. 1992), active stars with high flare activity
(Mathioudakis et al. 1995),
 or the widespread
presence of lithium in very cool dwarfs (Favata et al. 1996).

The  Li~{\sc i} EW variations that we observe are clearly correlated
with the temporal evolution of the flare (Fig.~2 and 4),
and large changes observed in the core of the Li~{\sc i}, 
as predict the models of Li production in flares (Livshits 1997).
Thus taking into account that
the other possible causes of variability have been eliminated above
{\it we suggest that this Li~{\sc i} is produced by spallation reactions
in the flare.}
This is the first time  that such Li~{\sc i} enhancement associate
with a stellar flare is reported, and probably the long-duration of
this flare  is a key factor for this detection.
The observed $^6$Li/$^7$Li ratio also support this hypothesis.

\vspace{0.4cm}
\hrule

\newpage
\hrule
\vspace{0.2cm}
\begin{center}
{\huge\it $^6$Li/$^7$Li ratio  enhancement}
\end{center}

Another signature of Li production from spallation reactions is
that the $^6$Li/$^7$Li isotopic ratio should increase.
The predicted $^6$Li/$^7$Li ratio for the lithium produced by spallation is
$\approx$~0.4 (Audouze 1970) or $\approx$~0.5 (Walker et al. 1985)
while the ratio measured in the Sun is
between 0.01 and 0.04 (M\"{u}ller et al. 1975),
in Population I stars is $\leq$~0.04
(Andersen et al. 1984; Maurice et al. 1984; Rebolo et al. 1986;
Pallavicini et al. 1987),
and in Population II stars is $\approx$~0.05
(Smith et al. 1993; Hobbs \& Thorburn 1994).

In order to estimate the $^6$Li/$^7$Li in our high resolution spectra
we adopt the method used by Herbig (1964) based on the shift of the center
of gravity (cog) of the Li~{\sc i} blend toward longer wavelengths as the
 $^6$Li fraction increases.
If each  Li~{\sc i} component is weighted by its {\it gf}-value,
pure $^7$Li would produce a cog wavelength of 6707.8117~\AA\, while
pure $^6$Li would be 6707.9713~\AA\, and, for weak lines, intermediate
mixtures would yield a wavelength, $\lambda_0$, between the two isotopes
that would be weighted by the $^6$Li/$^7$Li ratio,

\hspace{3.5cm}
$^6$Li/$^7$Li = ($\lambda_0$ - 6707.8117) / (6707.9713 - $\lambda_0$)

To determine the value of $\lambda_0$ we have measured the difference
between the Li~{\sc i} and Ca~{\sc i} features
($\Delta$ = Ca~{\sc i} - Li~{\sc i}) and adopt a wavelength of
6717.681 \AA\ for the Ca~{\sc i} line.
We give this values and the corresponding $^6$Li/$^7$Li ratio obtained
in Table~1.
In order to test the possible errors
we have also measured this difference, $\Delta$, for other photospheric
lines included in the same spectral order than the Li~{\sc i} feature.
In Fig.5  we plot $\Delta$ for the Li~{\sc i} lines and the other lines.
The other line $\Delta$'s do not show any trend during the
observations and the $\sigma$ with respect to the mean value is $\approx$
0.008.
The Li~{\sc i} line $\Delta$ shows a tendency to decrease toward
the end of the flare, attaining a maximum difference 0.05.
This significant change in $\Delta$ and thus in the $^6$Li/$^7$Li ratio
is consistent with increasing $^6$Li during the flare.
This is what is predicted for the production of Li~{\sc i} by spallation
reactions.

\vspace{0.4cm}
\hrule

\newpage

\hrule
\vspace{0.5cm}
{\Large\bf References}
\scriptsize
\vspace{0.4cm}

--  Andersen J., Gustafsson B., Lambert D.L., 1984,  A\&A 136, 65

-- Audouze J., 1970, A\&A 8, 436

--  Barden, S. C. 1985, ApJ, 295, 162

-- Barrado D., 1996, Ph. D. Universidad Complutense de Madrid

--  Berdyugina S.V., Ilyin I., Tuominen I.,
1998a,  in:
{Cool Stars, Stellar Systems, and the Sun,
Tenth Cambridge Workshop}, J.A. Bookbinder \& R.A. Donahue, eds.,
ASP Conf. Ser. 155, San Francisco: ASP, CD-1477

-- Berdyugina S.V., Jankov S., Ilyin I., Tuominen I., Fekel F.C.,
1998b, A\&A 334, 863  

-- Boesgaard A. M., 1991, In: The Formation and Evolution 
of Stars Clusters, ASP Conf. Series 13, 463

-- Canal R. 1974, 
ApJ 189, 531 

-- Canal R., 
Isern J.,  Sanahuja B. 1975, 
ApJ 200, 646 

-- Chupp E.L., et al. 1973, Nature, 241,
333

--  Cutispoto, G., Pagano, I., \& Rodono, M. 1992, A\&A, 263, L3

--  Elvis M., Plummer D., Schachter J., Fabbiano G., 1992, ApJS 80,
257

-- Endl, M., Strassmeier, K.G., \& K\"urster, M. 1997, A\&A, 328, 565

-- Engvold O., Kjeldseth Moe O., Maltby P., 1970, A\&A 9, 79

-- Favata F., 
Micela G.,  Sciortino S., 1996, A\&A 311, 951

-- Fekel F.C., 1996, In: Stellar Surface Structure, IAU Symp 176,
Strassmeier K.G., Linsky J.L. (eds.), Kluweer Acad. Publ., p. 345

-- Fern\'andez M., Miranda L.F., 1998, A\&A 332, 629

-- Fowler 
W.A., Burbidge G.R.,  Burbidge E.M., 1955, 
ApJS 2, 167 

-- Giampapa M.S., 1984, ApJ 277, 235

-- Graffagnino, V.G., Wonnacott, D. \& Schaeidt, S. 1995,
MNRAS, 275, 129

-- Herbig G.H., 1964, ApJ 140, 702

-- Hobbs L.M., Thorburn J.A., 1994, ApJ 428, L25

-- Huenemoerder D.P., Ramsey L.W., 1987, ApJ 319, 392

-- Hussain G.A.J., Unruh Y.C., Collier Cameron A, 
1997, MNRAS 288, 343

-- Hultqvist L., 1974, Solar Phys. 34, 25

-- Hultqvist L., 1977, 
Solar Phys. 52, 101 

-- Jeffries R.D., Byrne P.B., Doyle J.G., Anders G.J., James D.J.,
Lanzafame A.C., 1994, MNRAS 270, 153

-- Jeffries R.D., Bertram D., Spurgeon B.R.,
1995, MNRAS 276, 397

-- Karpen J.T.  Worden 
S.P., 1979, A\&A 71, 92 

-- Kotov Y.D., Bogovalov 
S.V., Endalova O.V.,  Yoshimori M., 1996, 
ApJ 473, 514 

--  K\"urster M. \& Schmitt J.H.M.M. 1996, A\&A, 311, 211

-- Livshits M.A., 1997, Solar Phys. 173, 377

-- Livingston W., Poveda A., Wang Y., 
1997, Advances in the Physics of Sunspots, 
B. Schmieder, J.C. del Toro Iniesta, M, V\'azquez (eds.), 
ASP Conf. Ser., 118, 86 

-- Mason K.O., et al., 1995, MNRAS 274, 1194

-- Mathioudakis M., et al., 1998, A\&A 302, 422 

-- Maurice E., Spite F., Spite M., 1984, A\&A 132, 278

-- Montes D., Sanz-Forcada J., Fern\'{a}ndez-Figueroa M.J.,
Lorente R., 1996, A\&A 310, L29

--  Montes D., Fern\'{a}ndez-Figueroa M.J., De Castro E.,
Sanz-Forcada J., 1997, A\&AS 125, 263

--  Montes D.,  Saar S.H., Collier Cameron A., Unruh Y.C.,
1998, MNRAS (in press)

-- M\"{u}ller E.A., Peytremann E., de la Reza T., 1975, Solar Phys. 41, 53

-- Murphy R.J., Hua X. 
-M., Kozlovsky B.,  Ramaty R., 1990, 
ApJ 351, 299 

-- Neuh\"auser R., et al., 1998, A\&A  334, 873 

-- O'Neal D., Saar S.H., Neff J.E., 1996, ApJ 463, 766

-- Ottmann, R. \& Schmitt, J.H.M.M. 1994, A\&A, 283, 871

-- Pallavicini R., Cerruti-Sola M., Duncan D.K., 
1987, 174, 116

-- Pallavicini R., Randich S., Giampapa M.S., 
1992, A\&A 253, 185

-- Pallavicini R., Cutispoto G., Randich S., Gratton R., 
1993, A\&A 267, 145

-- Patterer R.J., Ramsey L.W., Huenemoerder D.P., Welty A.D., 
1993, AJ 105, 1519

-- Pounds K.A., et al. 1993, MNRAS 260, 77a

-- Pye J.P., et al. 1995, MNRAS 274, 1165

-- Rebolo R., Crivellari L., Castelli F., Foing B., Beckman J.E.,
1986, A\&A 166, 195

-- Ritzenhoff S., Schr\"{o}ter e.H., Schmidt W., 1997 A\&A 328, 695

-- Robinson R.D., Thompson K., Innis J.L., 
1986, Proc. Astron. Soc. Aust. 6, 500

-- Ryter C., Reeves H., Gradsztajn E., Audouze J., 
1970, A\&A 8, 389

--  Schachter J., 1996, ApJ 463, 747

-- Smith V.V., Lambert D.L., Nissen P.E., 1993, ApJ 408, 262

-- Soderblom D.R., et al., 
1993, AJ 106, 1059

-- Soderblom D.R., 1996, in: Cool Stars, Stellar Systems, 
and the Sun, R. Pallavicini \& A.K. Dupree (eds.) ASP Conf. Series 109, 315

-- Stuik R., Bruls J.H.M.J., Rutten R.J., 1997, A\&A 322, 911 

-- Walker T.P., Mathews G.J., Viola V.E., 1985, ApJ 299, 745

-- Yoshimori M.,  et al.,
1994, ApJ 90, 639

-- Zirin H., 1988, in Astrophysics of the Sun,
(Cambridge University Press)

\vspace{0.5cm}
\hrule

\newpage
\vspace{0.2cm}
\normalsize

\begin{table}
\caption{Measured Li~{\sc i} 6708\AA\ line parameters
 \label{tab:li}}
\begin{center}
\begin{tabular}{cccccc}
\hline
\hline
HJD          & EW & I &
$\Delta$ Ca - Li & $\lambda_0$ & $^6$Li/$^7$Li \\
(2450000+)   & \AA\ &   &  \AA\ &  \AA\ &   \\
\hline
              &       &     &  &  \\
 826.8794     & 0.102 & 0.146 & 9.892 & 6707.789 & 0. \\
 827.8537     & 0.122 & 0.149 & 9.899 & 6707.782 & 0. \\
 828.8925     & 0.133 & 0.161 & 9.898 & 6707.783 & 0. \\
 829.8929     & 0.126 & 0.165 & 9.886 & 6707.795 & 0. \\
 830.9239     & 0.136 & 0.164 & 9.889 & 6707.792 & 0. \\
 831.9308     & 0.138 & 0.177 & 9.880 & 6707.801 & 0. \\
 832.9098     & 0.153 & 0.178 & 9.861 & 6707.820 & 0.06 \\
 835.8891     & 0.130 & 0.144 & 9.853 & 6707.828 & 0.11 \\
              &       &     &  &  \\
\hline
\end{tabular}
\end{center}
\end{table}

\newpage

\begin{table}
\caption{Photospheric line parameters
 \label{tab:linesp}}
\begin{center}
\begin{tabular}{lcccccccccccccc}
\hline
\hline
Line Id. (M) & $\lambda$ & E    &
EW & $\sigma$ & EW$_{\rm max - min}$  \\
             & (\AA)     & (eV) &
(\AA) &       & (\AA)                   \\
\hline
Li~{\sc i} (1)   & 6707.8   & 0.000 & 0.130 & 0.014 & 0.051 \\
                 &          &       &        &       \\
Ca~{\sc i} (18)  & 6462.566 & 2.523 & 0.344 & 0.008 & 0.024 \\
Ca~{\sc i} (18)  & 6471.660 & 2.526 & 0.192 & 0.006 & 0.017 \\
Ca~{\sc i} (1)   & 6572.781 & 0.000 & 0.173 & 0.007 & 0.023 \\
Ca~{\sc i} (32)  & 6717.685 & 2.709 & 0.225 & 0.009 & 0.024 \\
                 &          &       &        &       \\
Fe~{\sc i} (268) & 6546.245 & 2.758 & 0.174 & 0.006 & 0.021 \\
Fe~{\sc i} (111) & 6663.446 & 2.424 & 0.180 & 0.005 & 0.016 \\
Fe~{\sc i} (268) & 6703.573 & 2.758 & 0.101 & 0.005 & 0.017 \\
Fe~{\sc i} (34)  & 6710.310 & 1.485 & 0.094 & 0.003 & 0.009 \\
                 &          &       &        &       \\
Ni~{\sc i} (43)  & 6643.641 & 1.676 & 0.172 & 0.005 & 0.019 \\
\hline
\end{tabular}
\end{center}
\end{table}


\pagestyle{empty}
\normalsize

\clearpage

\begin{figure*}
\vspace{-0.9cm}
{\psfig{figure=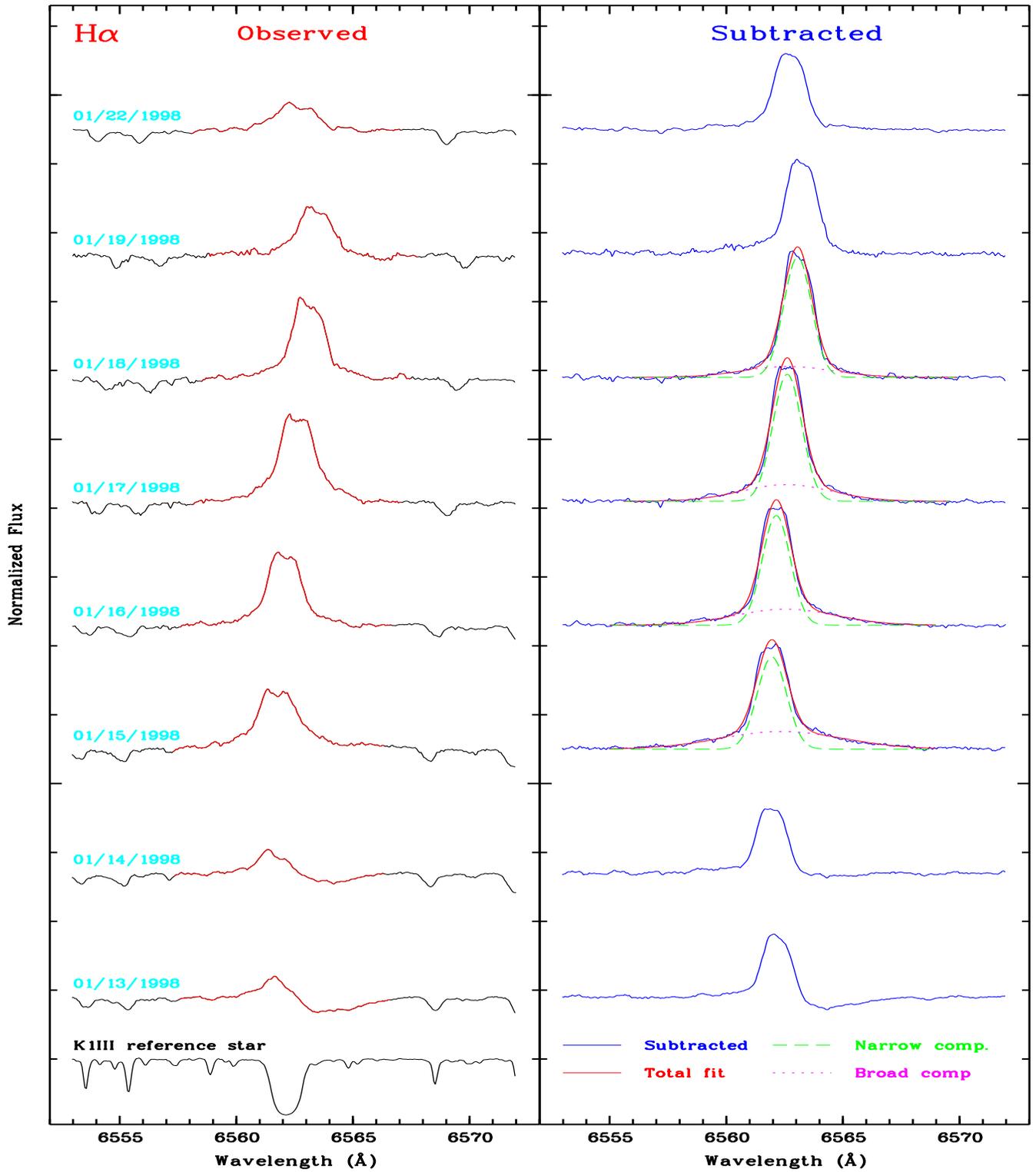,bbllx=45pt,bblly=30pt,bburx=516pt,bbury=786pt,height=20.8cm,width=18.0cm,clip=}}
\caption[ ]{H$\alpha$ observed spectra (left panel),
and after the spectral subtraction (right panel)
\label{fig:rej0743_mcd98_ha} }
\end{figure*}

\setcounter{figure}{0}
\begin{figure*}
{\psfig{figure=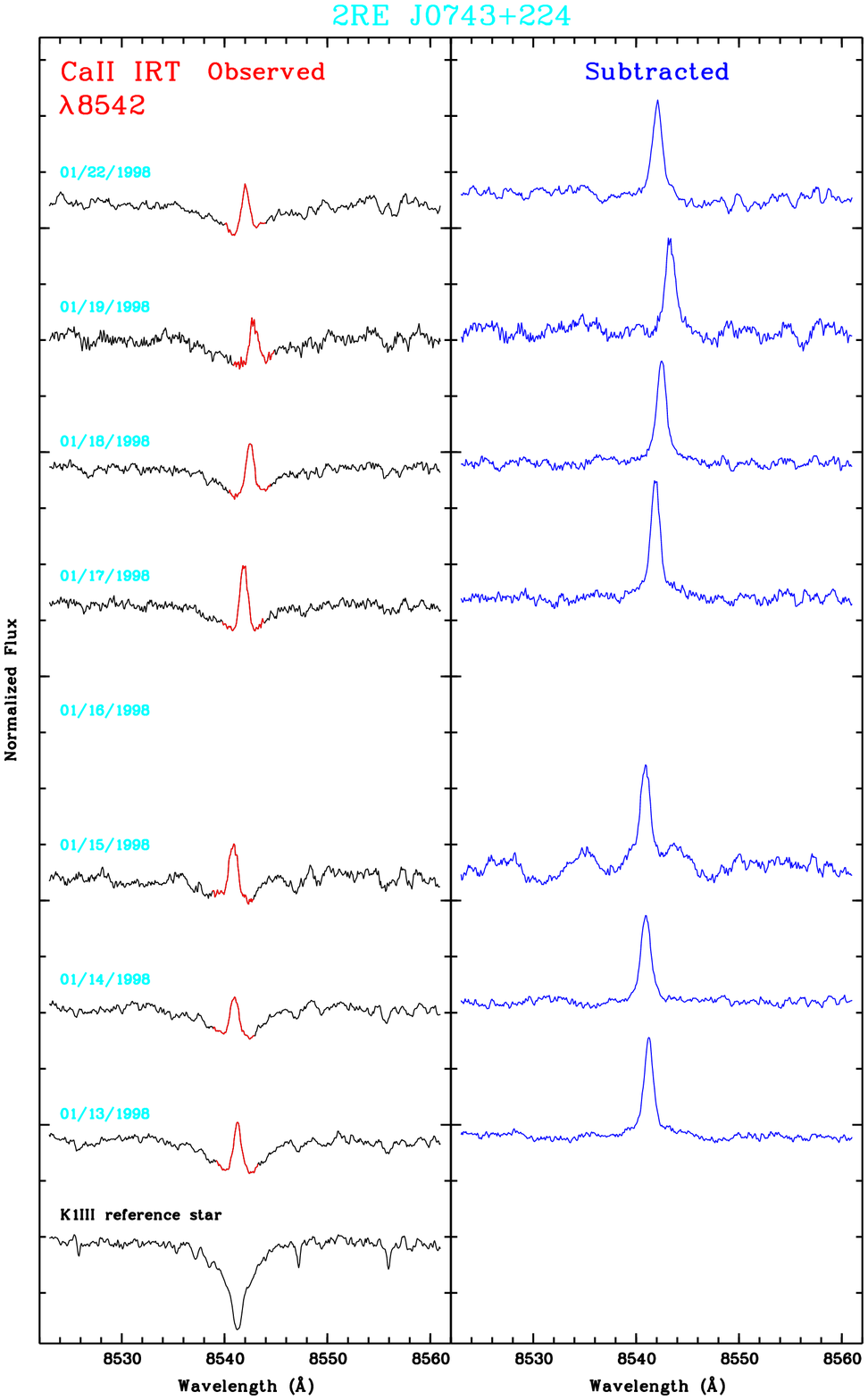,bbllx=45pt,bblly=30pt,bburx=516pt,bbury=786pt,height=20.8cm,width=18.0cm,clip=}}
\caption[ ]{Ca~{\sc ii} IRT $\lambda$$8542$ observed spectra (left panel),
and after the spectral subtraction (right panel)
\label{fig:rej0743_mcd98_ca} }
\end{figure*}

\clearpage

\begin{figure*}
\begin{center}
{\psfig{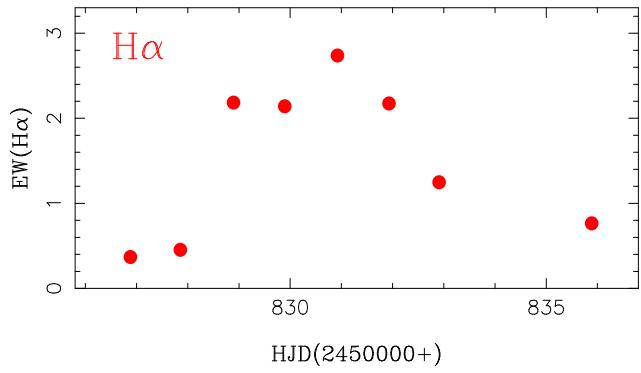}}
\caption[ ]{Temporal evolution of the H$\alpha$ EW during the flare
\label{fig:ha_ews} }
\end{center}
\end{figure*}

\clearpage

\begin{figure*}
\vspace{-0.9cm}
{\psfig{figure=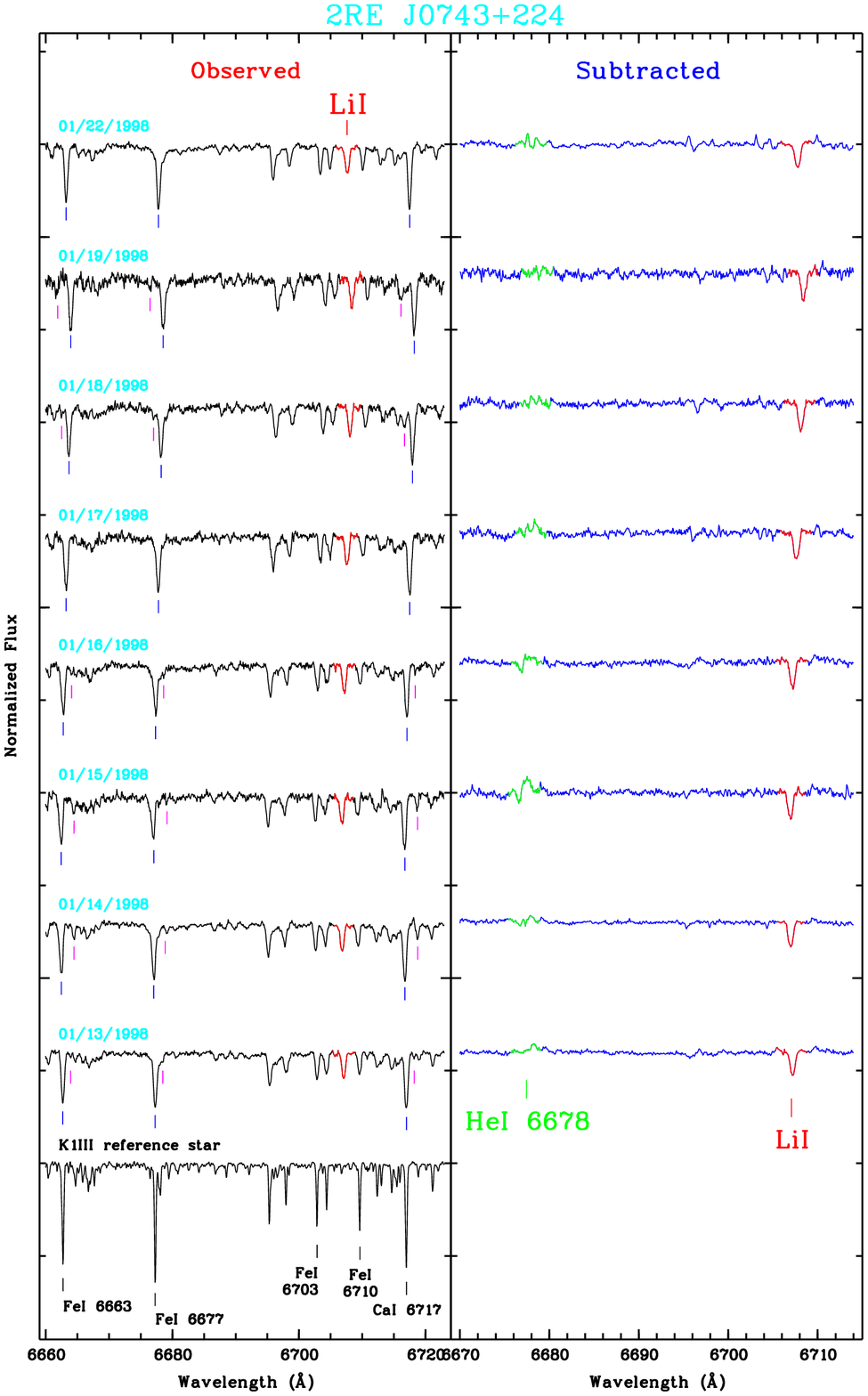,bbllx=45pt,bblly=30pt,bburx=516pt,bbury=786pt,height=20.8cm,width=18.0cm,clip=}}
\caption[ ]{Li~{\sc i} observed spectra (left panel),
and after the spectral subtraction (right panel)
\label{fig:rej0743_mcd98_ha} }
\end{figure*}

\clearpage

\begin{figure*}
{\psfig{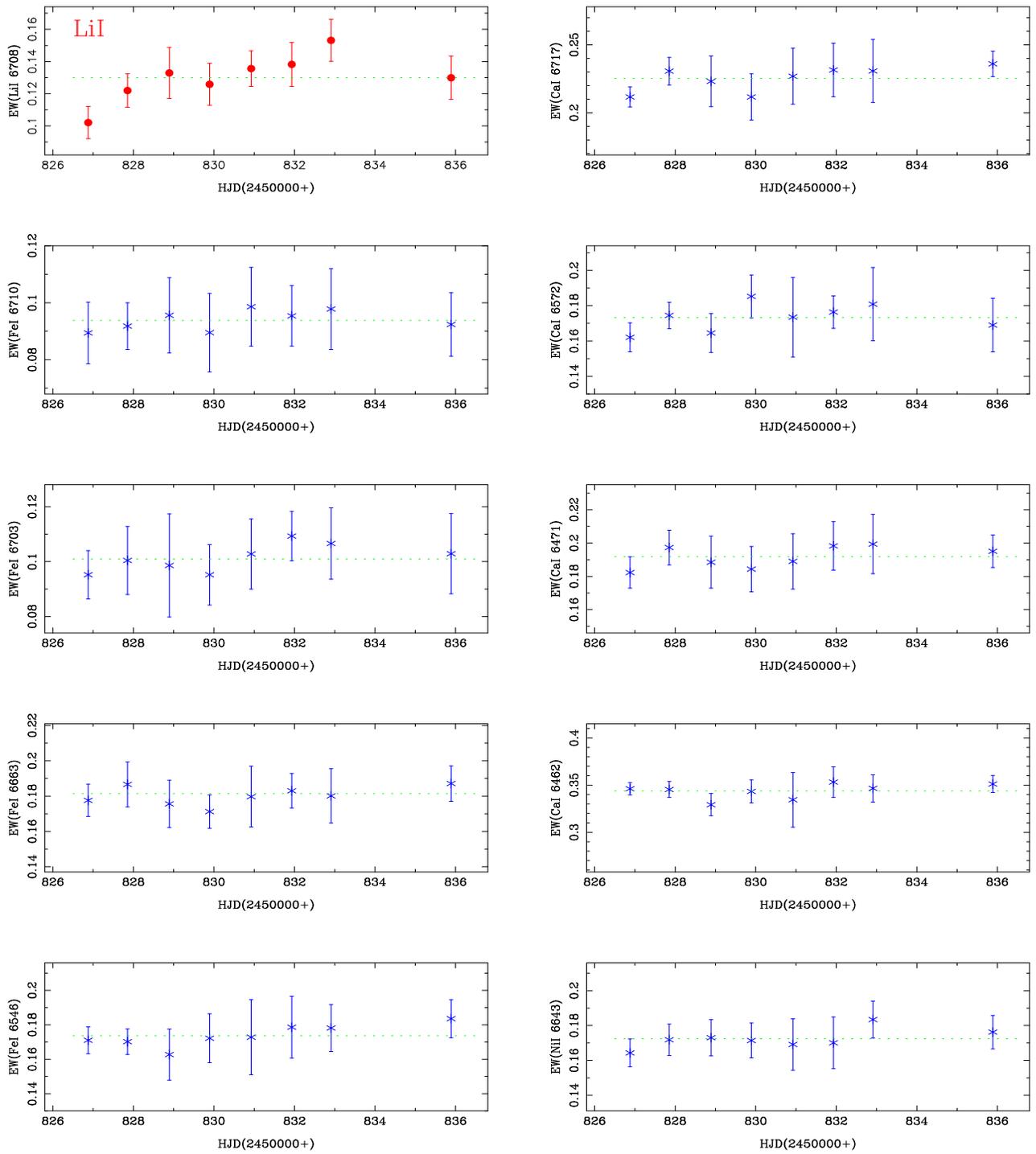}}
\caption[ ]{Measured EW for the Li~{\sc i} 6708\AA\ line and other
photospheric
lines
\label{fig:li_ews} }
\end{figure*}

\clearpage

\begin{figure}
{\psfig{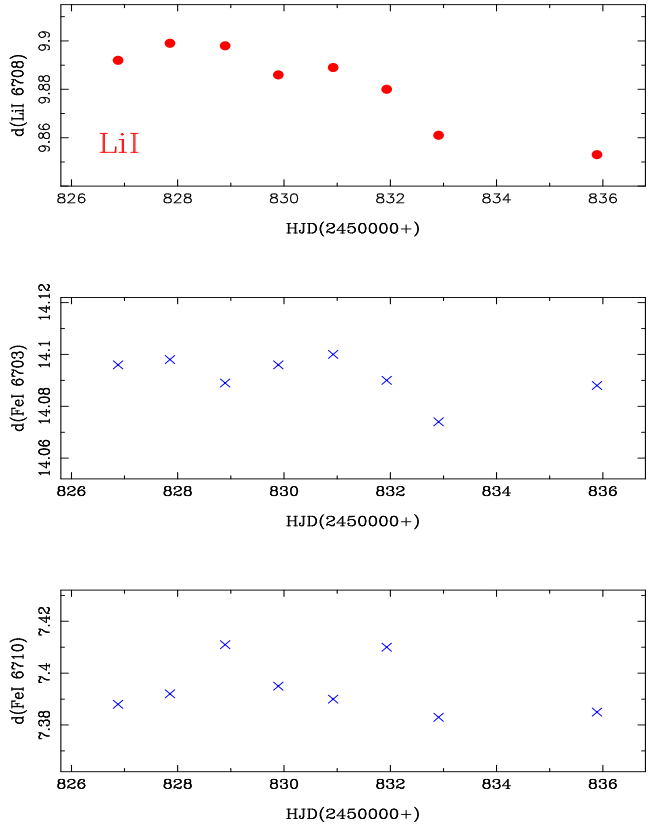}}
\caption[ ]{
Wavelength difference
between the Li~{\sc i} and Ca~{\sc i} features
as a measure of the $^6$Li/$^7$Li ratio
and the same difference for other photospheric lines
\label{fig:li_ll} }
\end{figure}

\end{document}